\documentclass[twocolumn,showpacs]{revtex4}

\begin{document}

\title{Quasi-equilibrium binary black hole sequences 
for \\ puncture data derived from helical Killing vector conditions}

\author{Wolfgang Tichy and Bernd Br\"ugmann}

\affiliation{
Center for Gravitational Physics and Geometry and 
Center for Gravitational Wave Physics\\
Penn State University, University Park, PA 16802}

\date{July 7, 2003}

\input epsf

\begin{abstract}
We construct a sequence of binary black hole puncture data derived 
under the assumptions 
(i) that the ADM mass of each puncture as measured in the
asymptotically flat space at the puncture stays constant along the
sequence, and (ii) that the orbits along the sequence are quasi-circular in
the sense that several necessary conditions for the existence of
a helical Killing vector are satisfied. These conditions
are equality of ADM and Komar mass at infinity and equality of the ADM
and a rescaled Komar mass at each puncture. In this paper we explicitly
give results for the case of an equal mass black hole binary
without spin, but our approach can also be applied in the general case.
We find that up to numerical accuracy the apparent horizon mass
also remains constant along the sequence and that the prediction for the
innermost stable circular orbit is similar to what has been found with the
effective potential method.
\end{abstract}

\pacs{
04.20.Ex    
04.25.Dm,   
04.30.Db,   
95.30.Sf    
%
}

\maketitle

\section{Introduction}

Binary black hole inspirals and mergers are promising sources for
ground-based interferometric gravitational wave detectors such as
GEO600, LIGO and TAMA~\cite{Schutz99}.  These systems are highly
relativistic once they enter the sensitive frequency band
of the detector. Hence, in order to predict the gravitational waves from
these sources numerical simulations will be needed, which in turn
require astrophysically realistic initial data.
Several methods to produce initial data for binary black hole systems exist
(see~\cite{Cook:2000} for a recent review). It is, however, not
clear yet which method should be used to obtain initial data
that are as realistic as possible in terms of astrophysical content.
During the inspiral, Post-Newtonian (PN) theory predicts that
the two black holes will be in quasi-circular orbits around each other 
with a radius which shrinks on a timescale much larger than the 
orbital timescale. One approach to incorporate this PN information
is to directly use the PN metric when solving the constraint equations to 
obtain initial data \cite{Tichy02}. 
Another approach is the so called effective potential method which
identifies circular orbits with minima in a suitably defined binding energy
\cite{Cook94}.
Yet another approach is to concentrate 
on the fact that quasi-circular inspiral orbits evolve slowly, which
means that the initial data should have an approximate helical Killing 
vector \cite{Gourgoulhon:2001ec,Tichy03a}.

Not only are there different methods to define quasi-circularity, but
there are different ways to construct binary black hole initial data
on arbitrary orbits.
Due to its simplicity, the puncture construction of black hole data and the puncture
evolution method \cite{Brandt97b,Bruegmann97} have been used in many
black hole simulations, in fact to date all gravitational wave forms
obtained numerically for binary black hole inspirals are based on puncture
initial data, see e.g.\
\cite{Alcubierre00b,Alcubierre02a,Baker00b,Baker:2001sf,Baker:2001nu,Baker:2002qf}.
Although there are several alternatives~\cite{Cook:2000}, we
concentrate on puncture data in this paper.

\pagebreak

The first quasi-circular puncture sequence was obtained
by Baumgarte~\cite{Baumgarte00a} using the effective potential method and
assuming constant apparent horizon mass along the
sequence. Baker~\cite{Baker02a} derived a puncture sequence based on
a variational principle that leads to an effective potential method
with constancy of the Arnowitt-Deser-Misner (ADM) mass at the puncture.
Furthermore, within a certain error a puncture sequence can also be
obtained~\cite{Baker:2002qf} by using the orbital parameters computed
for Misner type black hole excision data by Cook~\cite{Cook94}. 

Recently, we have shown how the approximate helical Killing vector
idea can be applied to puncture data \cite{Tichy03a}. We have found
that puncture data with orbital parameters determined using the
effective potential method approximately fulfill several necessary
conditions for the existence of a helical Killing vector, namely the
equality of the ADM and Komar mass at infinity, and equality of ADM
mass and a suitably scaled Komar mass at both punctures.

The main result of the present paper is that this work can be extended
to obtain sequences of black hole puncture data in quasi-circular
orbits. We use the same necessary conditions derived for an
approximate helical Killing vector plus the assumption that the ADM
mass of each puncture as measured in the asymptotically flat space at
the puncture stays constant. Our results for the innermost stable
circular orbit (ISCO) are close to what
\cite{Cook94} and \cite{Baumgarte00a} find with the
effective potential method. We also find that up to numerical accuracy
the apparent horizon mass stays constant along our sequence, which
is usually assumed to hold in conjunction with the effective potential
method. Furthermore, we provide detailed information for the orbital
parameters of the puncture sequence, which has not yet been available
in the literature since~\cite{Baumgarte00a,Baker02a} focus on the ISCO
and do not provide this information.


\section{Quasi-circular sequences}

The orbits of two puncture black holes are completely described by 
the ADM mass $M^{ADM}_A$ and spin $S^i_A$ of each black hole
computed in the asymptotically flat space at the puncture,
the coordinate distance $D$ between the two black holes, and
the the momentum parameter $P^i_A$ of each black hole
($A = 1,2$ for the two black holes). 
If we choose coordinates such that the total ADM momentum is zero, the two
black holes must have equal and opposite momenta, and hence only one momentum
parameter $P=|P^i_A|$ is needed. Thus the system is characterized by 
the 10 physical parameters $M^{ADM}_A$, $S^i_A$, $P$, and $D$. 
In addition, we have the 2 parameters $c_A$ 
which fix the value of the lapse $\alpha$ at each puncture. 
In our approach $\alpha$ is needed to compute Komar mass integrals
at both punctures and at infinity, and we determine $\alpha$
from a maximal slicing equation \cite{Tichy03a}. Hence, puncture data
together with the lapse are described by 12 parameters.

We want to construct a sequence of
quasi-circular orbits, i.e.\ we want to find $M^{ADM}_A$, $S^i_A$, $P$, and
$c_A$ such that for any $D$ we have the same two black holes on a
quasi-circular orbit. In order to assure that we are always dealing with the
same two black holes, we impose that
\begin{eqnarray}
\label{M_A_const}
M^{ADM}_A &=& const ,
\end{eqnarray}
\begin{eqnarray}
\label{S_A_const}
S^i_A  &=& const .
\end{eqnarray}
Note that this choice is not unique, since individual masses and
spins of black holes in binaries contain ambiguities. But it is expected
that $M^{ADM}_A$ and $S_A$ will be approximately constant during the
inspiral of realistic black holes, at least in the 
quasi-stationary regime \cite{Blackburn92}. 
To assume constancy of the masses at the punctures was also 
considered by Baker \cite{Baker02a}.
In addition, to ensure quasi-circular orbits, we impose the three 
necessary helical Killing vector conditions \cite{Tichy03a}
\begin{equation}
\label{MK_MADM}
I_K({\alpha},S_{\infty}) = M^{ADM}_{\infty}
\end{equation}
and
\begin{equation}
\label{MK_MADM1}
I_K({\alpha},S_{A}) = c_A M^{ADM}_{A} .
\end{equation}
Here
\begin{equation}
\label{Komar2}
I_K({\alpha},S)= \frac{1}{4\pi}\oint_{S}\bar{\nabla}_i{\alpha} \ d\bar{S}^i
\end{equation}
is the Komar mass, evaluated either on a sphere $S_{\infty}$ at infinity 
or on infinitesimal spheres $S_{A}$ around each puncture.
Now the 11 parameters $M^{ADM}_A$, $S^i_A$, $P$, and $c_A$ can be determined
from the 11 conditions (\ref{M_A_const}), (\ref{S_A_const}), (\ref{MK_MADM}),
and (\ref{MK_MADM1}) for any given black hole separation $D$.

For simplicity, we now restrict ourselves to non-spinning equal mass
binaries with $M^{ADM}_A=M/2$. 
In this case $c_1=c_2=c$ will hold, so that we are left with only
the 3 parameters $M$, $P$, and $c$, which can be fixed using the 3 conditions
(\ref{M_A_const}), (\ref{MK_MADM}), and (\ref{MK_MADM1}).
The number of parameters can be further reduced through 
rescaling by $M$. When this is done it is sufficient to
find the 2 parameters $P/M$ and $c$ which satisfy
conditions (\ref{MK_MADM}) and (\ref{MK_MADM1}) for each given $D/M$.
Determining the appropriate parameters involves root finding, which is
computationally expensive since in each iteration two elliptic
equations for the conformal factor $\phi$ and for the lapse $\alpha$
have to be solved.

In order to construct sequences of quasi-circular orbits as described above,
we use the same numerical techniques as in \cite{Tichy03a}
combined with a Newton-Raphson method to find the parameters $P/M$ and $c$.
All numerical grids have a uniform resolution of $h=0.0625 M$, and
the outer boundary is located at $R_1=12 M$. To obtain accurate values for
the ADM and Komar integrals needed in Eq.~(\ref{MK_MADM}) we use
volume integrals over the numerical grid (out to $R_1=12 M$) 
plus a correction from integrating outside the grid out to 
a larger distance $R_2$ (see appendix \ref{MassesAtInf}). 
To sufficiently reduce the error with this correction we 
set $R_2 = 10 R_1^2 /M = 1440 M$.
Then the masses at infinity have errors of only $0.0005\%$.
The ADM and Komar integrals at the punctures 
needed in condition (\ref{MK_MADM1})
are obtained by fourth order interpolation of the conformal factor and the 
lapse onto the location of the punctures. 
The error in the masses at the punctures is of order $0.03\%$,
which is thus also the error to within which we can determine our sequence,
as it is much larger than the error in the masses at infinity.

\begin{figure}
\epsfxsize=8.5cm 
\epsfbox{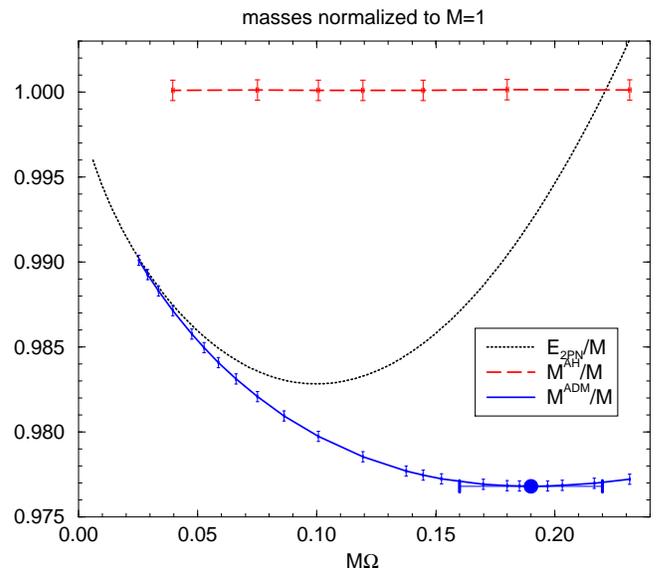}
\vspace{.4cm}
\caption{The ADM mass at infinity, the total 2PN energy, 
and the apparent horizon mass as a function of angular velocity. 
The minimum in $M^{ADM}_{\infty}$ at $M \Omega_{min}=0.19\pm 0.03$
is marked by a dot.
}
\label{Mass_plot}
\end{figure}
In Fig.~\ref{Mass_plot} we show our results for the 
ADM mass $M^{ADM}_{\infty}$ at infinity and the apparent horizon mass 
\begin{equation}
\label{M_AH}
M^{AH} = \sqrt{A_1/16\pi} + \sqrt{A_2/16\pi}. 
\end{equation}
Here $A_1$ and $A_2$ are the areas of the two apparent horizons.
In order to obtain coordinate independent results, all our graphs are plotted
versus the angular velocity \cite{Tichy03a}
\begin{equation}
\label{OmegaFromI_K}
\Omega = 
\frac{M^{ADM}_{\infty} - I_K({\alpha},S_1) - I_K({\alpha},S_2)}
{2J^{ADM}_{\infty}} ,
\end{equation}
instead of the coordinate separation $D$. 
For comparison, we also show the 2PN total energy \cite{Tichy02,Schaefer93}
\begin{eqnarray}
\label{PN_Etot}
E_{2PN}&=& M-\frac{\mu M}{2D}
\Bigg[1 + \Big(\frac{\mu}{M}-7\Big)\frac{M}{4D} \nonumber \\
&+&\!\!\! \Bigg(\frac{\mu^2}{M^2}+20\frac{\mu}{M}-9\Bigg)
                     \frac{M^2}{8D^2} \Bigg]
\end{eqnarray}
versus the 2PN angular velocity \cite{Tichy02,Schaefer93}
\begin{eqnarray}
\label{ang_vel}
\left( M \Omega_{2PN} \right)^2&=&  \frac{64(D/M)^3}{(1+2D/M)^6}
+ \frac{\mu}{M} \left( \frac{M}{D} \right)^4  \nonumber \\
&+&\!\!\! \left(\frac{\mu^2}{M^2} -\frac{5}{8}\frac{\mu}{M}\right) 
    \left( \frac{M}{D} \right)^5 .
\end{eqnarray}
Note that $\Omega_{2PN}$ in Eq.~(\ref{ang_vel}) is written such that
$\Omega_{2PN}$ is exact up to all PN orders in the limit of $\mu/M
\rightarrow 0$, while for $\mu/M > 0$ Eq.~(\ref{ang_vel}) is accurate up to
2PN order. It should be kept in mind, however, that 
both, the resummation of Eq.~(\ref{ang_vel}), as well as the fact
that $E_{2PN}$ is given as a function of $D$ instead of say $\Omega_{2PN}$,
will change the $E_{2PN}$ versus $\Omega$ curve 
by PN terms higher than the $2PN$ terms included. This effect becomes
noticeable around $M \Omega \gtrsim 0.1$, where these 
higher order terms become important.

The ADM mass $M^{ADM}_{\infty}$ approaches the PN result for large
separations (i.e. small $\Omega$) as expected. Also, within our accuracy
$M^{AH}$ is constant along the sequence. In previous
works \cite{Cook94,Baumgarte00a} this property was usually
enforced in place of Eq.~(\ref{M_A_const}). It is interesting that 
the necessary conditions (\ref{MK_MADM}) and (\ref{MK_MADM1}) for the
existence of a helical Killing vector imply that 
$M^{ADM}_A= const$ and $M^{AH}=const$ are equivalent, and that in fact
$2M^{ADM}_A=M=M^{AH}$. 
The ADM mass $M^{ADM}_{\infty}$ has a minimum at 
$M \Omega= M \Omega_{min}=0.19\pm 0.03$,
which is usually interpreted as the innermost stable circular orbit (ISCO).
This ISCO minimum occurs (within error bars) at the same angular velocity as
the ones found by Cook \cite{Cook94} and Baumgarte \cite{Baumgarte00a},
using the effective potential method. Hence, results obtained using the
assumption of a helical Killing vector are very similar to results found
using the effective potential method.
The reason for the relatively large error bar in $M \Omega_{min}$ comes from
the fact that the minimum in $M^{ADM}_{\infty}$ is very shallow, so that 
the errors of order $0.03\%$ in our mass determination get amplified.
Note, however, that our error estimate for $M \Omega_{min}$ is very 
conservative, as we have assumed that $M^{ADM}_{\infty}$ might randomly vary
within $0.03\%$. If we instead assume that all values for 
$M^{ADM}_{\infty}$ are systematically too large or
too small by $0.03\%$, the errorbar would be much smaller.

\begin{figure}
\epsfxsize=8.5cm 
\epsfbox{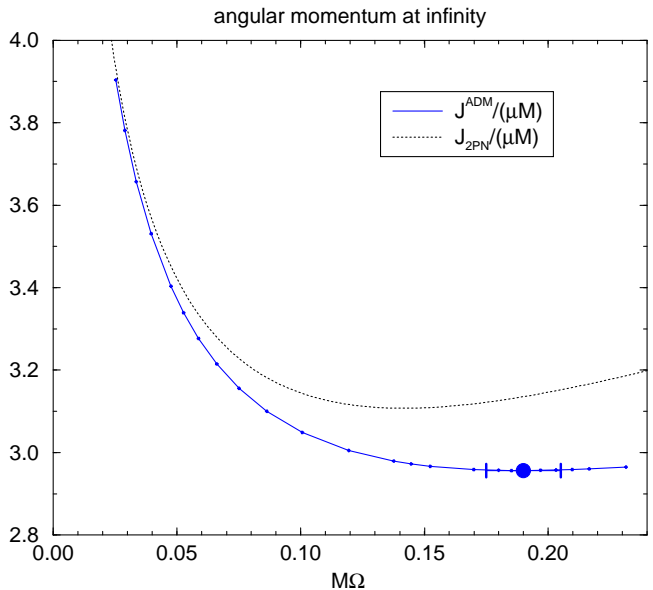}
\vspace{.4cm}
\caption{The ADM angular momentum at infinity and the 2PN angular momentum
as a function of angular velocity.
The minimum in $J^{ADM}_{\infty}$ at $M \Omega_{min}=0.190\pm 0.015$
is marked by a dot.}
\label{J_JPN}
\end{figure}

In Fig.\ \ref{J_JPN} we show the ADM angular momentum $J^{ADM}_{\infty}$ 
as well as the 2PN angular momentum \cite{Tichy02,Schaefer93}
\begin{equation}
\label{PN_J}
J_{2PN}^2 =  \mu^2 M D \Bigg[1 + 4\frac{M}{D}  
  +  \frac{\left(74 - 43 \frac{\mu}{M}\right) M^2}{8D^2} \Bigg]
\end{equation}
versus $M\Omega$, which agree for small $\Omega$ as expected. 
As we can see, $J^{ADM}_{\infty}$ has its minimum at the same 
$\Omega$ as $M^{ADM}_{\infty}$. However, as 
$J^{ADM}_{\infty}$ varies more strongly than $M^{ADM}_{\infty}$
the minimum is less shallow, which allows us to locate it with 
better accuracy, with the result $M \Omega= M \Omega_{min}=0.190\pm 0.015$.
Note that even though the error in $M \Omega_{min}$ is relatively large,
we are able to determine all other quantities with an error of only about
$0.03\%$.

Next, we list fits of several quantities
along our numerically computed sequence. All fitted quantities are 
given as functions of the dimensionless orbital radius
\begin{equation}
x=\frac{D}{2M} ,
\end{equation}
for the case of an equal mass binary. We find
\begin{eqnarray}
\frac{m}{M} &=& 
  \frac{1}{2} - \frac{0.0668549}{x} - \frac{0.0280090}{x^{\frac{3}{2}}}
  + \frac{0.0906937}{x^2}   \nonumber \\
&&			    - \frac{0.139304}{x^{\frac{5}{2}}}
  + \frac{0.0973014}{x^3}   - \frac{0.0238708}{x^{\frac{7}{2}}}
,\\
\frac{P}{M} &=&   
    \frac{1}{4{\sqrt{2x}}} - \frac{0.0123341}{x}
  + \frac{0.274250}{x^{\frac{3}{2}}} - \frac{0.317647}{x^2} 
 \nonumber \\
&&			 + \frac{0.554104}{x^{\frac{5}{2}}}
  - \frac{0.407033}{x^3} + \frac{0.101471}{x^{\frac{7}{2}}}
,\\  
c &=&  
    1                       + \frac{0.0243978}{{\sqrt{x}}}
      - \frac{0.509913}{x}  + \frac{0.629042}{x^{\frac{3}{2}}} \nonumber \\
&&    - \frac{1.00585}{x^2} + \frac{0.727406}{x^{\frac{5}{2}}}
      - \frac{0.177440}{x^3}
,\\
\frac{M^{ADM}_{\infty}}{M} &=&   
    1 - \frac{1}{16x}      + \frac{0.000684336}{x^{\frac{3}{2}}}
      + \frac{0.0499293}{x^2} \nonumber \\
&&			      - \frac{0.00649266}{x^{\frac{5}{2}}}
      + \frac{0.0304472}{x^3} - \frac{0.0561704}{x^{\frac{7}{2}}} \nonumber \\
&&    + \frac{0.0209257}{x^4}
,\\
\frac{J^{ADM}_{\infty}}{\mu M} &=&  2x \frac{P}{ \mu } = 8x \frac{P}{ M } ,
\\
M \Omega &=&   
    \frac{1}{2\,{\sqrt{2}}\,x^{\frac{3}{2}}} + \frac{0.0158006}{x^2}
  - \frac{0.347617}{x^{\frac{5}{2}}}         + \frac{0.275492}{x^3}
\nonumber \\
&&- \frac{0.175001}{x^{\frac{7}{2}}}         +\frac{0.0969903}{x^4}
  - \frac{0.0236081}{x^{\frac{9}{2}}}      
.\end{eqnarray}
For the given expansions the fitting error is
about $5\times 10^{-6}$. Our fits can be used to put two punctures into a
quasi-circular orbit for any desired separation $D$. For the reader's
convenience we also list the results of these fits for a variety of
separations in Tab.~\ref{Seq_table}.
\begin{table}
\caption{Parameters for quasi-circular orbits of binary black hole 
puncture data. The errors in the quantities listed are of order $0.03\%$.
\label{Seq_table}}
\begin{tabular}{|l|l|l|l|l|l|l|}
\hline
$D/2M$	& $m/M$	& $P/M$	& $c$	& 
$\frac{M^{ADM}_{\infty}}{M}$	& $\frac{J^{ADM}_{\infty}}{\mu M}$ & $M \Omega$ \\
\hline
 1.0300 & 0.43193 & 0.35879 & 0.69584 & 0.97681 & 2.9564 & 0.19008 \\ 
 1.0500 & 0.43318 & 0.35196 & 0.70109 & 0.97682 & 2.9564 & 0.18653 \\ 
 1.0750 & 0.43469 & 0.34380 & 0.70743 & 0.97683 & 2.9567 & 0.18223 \\ 
 1.1000 & 0.43613 & 0.33604 & 0.71354 & 0.97685 & 2.9571 & 0.17809 \\ 
 1.1250 & 0.43752 & 0.32865 & 0.71942 & 0.97689 & 2.9578 & 0.17409 \\ 
 1.1500 & 0.43885 & 0.32160 & 0.72509 & 0.97693 & 2.9587 & 0.17024 \\ 
 1.1750 & 0.44012 & 0.31488 & 0.73055 & 0.97698 & 2.9599 & 0.16652 \\ 
 1.2000 & 0.44135 & 0.30846 & 0.73581 & 0.97703 & 2.9612 & 0.16292 \\ 
 1.2250 & 0.44253 & 0.30232 & 0.74089 & 0.97710 & 2.9627 & 0.15945 \\ 
 1.2500 & 0.44366 & 0.29645 & 0.74580 & 0.97717 & 2.9645 & 0.15610 \\ 
 1.3000 & 0.44581 & 0.28543 & 0.75511 & 0.97732 & 2.9684 & 0.14972 \\ 
 1.3500 & 0.44780 & 0.27529 & 0.76380 & 0.97750 & 2.9731 & 0.14375 \\ 
 1.4000 & 0.44965 & 0.26594 & 0.77194 & 0.97769 & 2.9785 & 0.13815 \\ 
 1.4500 & 0.45138 & 0.25728 & 0.77957 & 0.97789 & 2.9844 & 0.13290 \\ 
 1.5000 & 0.45300 & 0.24924 & 0.78673 & 0.97811 & 2.9909 & 0.12797 \\ 
 1.6000 & 0.45593 & 0.23480 & 0.79981 & 0.97856 & 3.0055 & 0.11895 \\ 
 1.7000 & 0.45853 & 0.22219 & 0.81144 & 0.97903 & 3.0217 & 0.11092 \\ 
 1.8000 & 0.46084 & 0.21108 & 0.82184 & 0.97951 & 3.0395 & 0.10375 \\ 
 1.9000 & 0.46291 & 0.20123 & 0.83120 & 0.98000 & 3.0586 & 0.097299 \\ 
 2.0000 & 0.46477 & 0.19243 & 0.83964 & 0.98048 & 3.0788 & 0.091487 \\ 
 2.2500 & 0.46870 & 0.17406 & 0.85755 & 0.98164 & 3.1330 & 0.079215 \\ 
 2.5000 & 0.47185 & 0.15956 & 0.87192 & 0.98272 & 3.1911 & 0.069443 \\ 
 2.7500 & 0.47442 & 0.14780 & 0.88369 & 0.98372 & 3.2517 & 0.061518 \\ 
 3.0000 & 0.47656 & 0.13808 & 0.89350 & 0.98461 & 3.3138 & 0.054988 \\ 
 3.5000 & 0.47992 & 0.12287 & 0.90891 & 0.98619 & 3.4404 & 0.044923 \\ 
 4.0000 & 0.48243 & 0.11148 & 0.92044 & 0.98750 & 3.5674 & 0.037589 \\ 
 4.5000 & 0.48439 & 0.10259 & 0.92940 & 0.98859 & 3.6934 & 0.032052 \\ 
 5.0000 & 0.48595 & 0.095433 & 0.93655 & 0.98952 & 3.8173 & 0.027753 \\ 
 6.0000 & 0.48830 & 0.084541 & 0.94726 & 0.99099 & 4.0580 & 0.021565 \\ 
 7.0000 & 0.48997 & 0.076578 & 0.95491 & 0.99211 & 4.2884 & 0.017378 \\ 
 8.0000 & 0.49123 & 0.070451 & 0.96064 & 0.99299 & 4.5089 & 0.014390 \\ 
 9.0000 & 0.49220 & 0.065559 & 0.96511 & 0.99369 & 4.7203 & 0.012171 \\ 
 10.000 & 0.49299 & 0.061542 & 0.96868 & 0.99427 & 4.9233 & 0.010468 \\ 
\hline
\end{tabular}
\end{table}

\section{Discussion}
\label{Discussion}

We have constructed a sequence of binary black hole puncture data 
by using the necessary helical Killing vector 
conditions (\ref{MK_MADM}) and (\ref{MK_MADM1})
and by assuming that $M^{ADM}_A$ is constant along the sequence. 
The numerical results are obtained for equal mass binaries without spin,
but our approach can also be applied in the general case.
Our sequence is close to quasi-equilibrium in the following sense.  
The time derivative of the trace of the extrinsic curvature is zero,
due to the choice of a maximal slicing lapse.
The conformal metric and conformal factor evolve on a timescale long
compared to the orbital timescale,
if we also compute a shift as in \cite{Tichy03a}, which removes the
longitudinal piece of the time derivative of the conformal metric.
With this gauge choice the time derivative 
of the tracefree part of the extrinsic curvature is also reduced, 
but it still evolves on the orbital timescale.

We find that the apparent horizon mass $M_{AH}$ is constant along the
sequence up to our numerical accuracy, 
and that $M_{AH}=M^{ADM}_1+M^{ADM}_2$.
Furthermore, the ISCO minimum in both $M^{ADM}_{\infty}$ and
$J^{ADM}_{\infty}$ occurs at the same $\Omega$, but the error bars on the
minimum on in $J^{ADM}_{\infty}$ are tighter. Previous results 
for the ISCO \cite{Cook94,Baumgarte00a} agree
with our result within error bars, but our ISCO is at a somewhat 
higher $\Omega$.

Baker~\cite{Baker02a} has constructed a puncture sequence 
using an effective potential method and the assumption
that $M^{ADM}_A$ is constant along the sequence. 
His results are similar to ours. Yet even though he has used fixed
mesh refinement for computational efficiency, his values seem to be less
accurate. In~\cite{Baker02a}, the outer boundary was put at large
distances in order to compute $M^{ADM}_{\infty}$ with sufficient
accuracy, but based on our numerical results the resolution near the
black holes appears to be rather low.

The parameters for quasi-circular orbits in Tab.~\ref{Seq_table}
can be compared with the data for the effective
potential method in the form given in \cite{Baker:2002qf},
which provides a sequence translated
from Misner type black hole excision data of Cook \cite{Cook93,Cook94}
to puncture data assuming that both types of data are numerically close.
Our data is likely to be more accurate than effective potential method
data since we do not have to 
find turning points in potentially very flat ADM mass curves
to determine if an orbit is circular.
Also, our masses are computed with much higher accuracy than in
\cite{Baker:2002qf}. For example, for 
$D/2=1.7846 M=1.8490 M^{ADM}_{\infty}$ 
there is a discrepancy of $1\%$. 
Yet part of the difference may be due to the fact 
that we use the helical Killing vector conditions 
(\ref{MK_MADM}) and (\ref{MK_MADM1})
instead of effective potential method to find quasi-circular orbits.
The fact that we have kept $M^{ADM}_A$ constant along the sequence and not
$M_{AH}$ as in \cite{Cook94} and \cite{Baumgarte00a} should not play 
a role as both seem to be equivalent along our sequence. 

Finally, we want to mention that G. Cook has informed us that
M. Hannam has just completed his Ph.D. thesis which includes the discussion
of a puncture sequence that shares some of the features of our work.

\begin{acknowledgments}
It is a pleasure to thank Pablo Laguna for discussions. We acknowledge the
support of the Center for Gravitational Wave Physics funded by the
National Science Foundation under Cooperative Agreement
PHY-01-14375. This work was also supported by NSF grant PHY-02-18750.
\end{acknowledgments}

\appendix

\section{Computing masses at infinity on a finite grid}
\label{MassesAtInf}

\begin{figure}
\epsfxsize=8.5cm 
\epsfbox{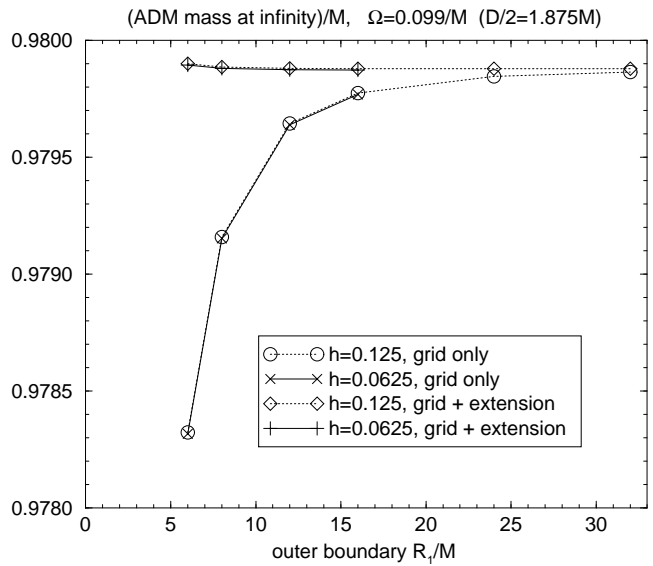}
\vspace{.4cm}
\caption{$M^{ADM}_{\infty}$ computed using integral (\ref{VolInt_approx0})
over the numerical grid only, 
compared with $M^{ADM}_{\infty}$ computed using the extended volume 
integral (\ref{VolInt_approx2}) with $R_2 =10 R_1^2 /M$
for two different resolutions $h$. 
The extended integral (\ref{VolInt_approx2}) 
gives much more accurate results. 
}
\label{M_ADM_grid_extension}
\end{figure}

In order to implement Eqs.~(\ref{MK_MADM}) and (\ref{MK_MADM1}), 
we have to determine the masses with sufficient accuracy. 
In particular, we have to compute the ADM mass at infinity to a 
rather high accuracy of about $10^{-4}M$. 
Since in the case of punctures \cite{Brandt97b} the 3-metric 
is conformally flat with a conformal factor
\begin{equation}	 
\phi = 1 + \frac{m_1}{2r_1} + \frac{m_2}{2r_2} + u,
\end{equation}  	 
the ADM mass at infinity is given by
\begin{equation}
\label{M_ADM_infty}
M^{ADM}_{\infty} 
= -\frac{1}{2\pi} \int \nabla^2 \phi \ dV 
= m_1 + m_2 - \frac{1}{2\pi} \int  \nabla^2 u \ dV ,
\end{equation}
Here $u$ has to be determined from an elliptic equation of the 
form \cite{Tichy03a,Brandt97b}
\begin{equation}	 
\nabla^2 u = f(u) ,
\end{equation}  
with boundary condition 
\begin{equation}
\label{u_BC}
\lim_{r\rightarrow \infty} u = 0 .	
\end{equation}
As our numerical grid does not extend all the way to infinity,
we cannot directly compute the volume integral in Eq.~(\ref{M_ADM_infty}).
Using a coordinate transformation in the radial direction, e.g.\ of
the ``FishEye'' type~\cite{Baker00b}, helps but does not necessarily give
sufficiently accurate results because of finite difference errors in
the far region.
Instead, the last term of Eq.~(\ref{M_ADM_infty}) can be approximated by
the integral
\begin{equation}
\label{I_1}
I_1 = \int_{0}^{R_1} f(u) \ dV
\end{equation}   
over the numerical grid up to the outer boundary denoted by $R_1$.
Then
\begin{equation}
\label{VolInt_approx0}
\int \nabla^2 u \ dV =  I_1 + O(1/R_1) ,
\end{equation}
where $O(1/R_1)$ is the
error due to truncating the integral at a finite radius $R_1$. This error is
often too large as it falls off only like $1/R_1$ (see the curves
labeled ``grid only'' in Fig.~\ref{M_ADM_grid_extension}).
For this reason we want to improve the approximation by integrating over a
region larger than the grid, which extends to at least $R_2 \sim R_1^2$, so
that the error goes down by one additional power in $1/R_1$. 
I.e.\ we want to approximate the volume integral by
\begin{equation}
\label{VolInt_approx1}
\int \nabla^2 u \ dV =  I_1 + I_2 + O(1/R_2) =  I_1 + I_2 + O(1/R_1^2) ,
\end{equation}
where
\begin{equation}
\label{I_2}
I_2 =   \int_{R_1}^{R_2} f(u) \ dV .
\end{equation}
The problem with this approach is that we have computed $u$ only on the
numerical grid, so that we have to somehow approximate $u$ outside the grid.
This can be achieved by noting that because 
of boundary condition (\ref{u_BC})
\begin{equation}
u = \frac{a_1}{r} + O(1/r^2)
\end{equation}
for large distances $r$, so that
\begin{equation}
I_1 = \int_{0}^{R_1} \nabla^2 u \ dV= \oint_{R_1}  \nabla_i u \ dS^i
    = -4\pi a_1 + O(1/R_1) .
\end{equation}
Hence outside the numerical grid $u$ can be approximated by 
\begin{equation}
\label{u_approx}
u = -\frac{1}{4\pi}\frac{I_1}{r} 
    + O\left(\frac{1}{r R_1}\right) + O\left(\frac{1}{r^2}\right)
\end{equation}
where $I_1$ is computed only on the grid. Also, note that for large $r$
\begin{equation}
\label{f_falloff}
f(u) \sim \frac{1}{r^4} ,  
\end{equation}
otherwise the ADM mass integral in Eq.~(\ref{M_ADM_infty}) would be
infinite.
Thus, if we insert Eq.~(\ref{u_approx}) into Eq.~(\ref{I_2}) and also
take into account the falloff behavior (\ref{f_falloff}) of $f(u)$
we arrive at
\begin{equation}
\label{I_2_approx}
I_2 =  \int_{R_1}^{R_2} f\left( -\frac{1}{4\pi}\frac{I_1}{r}  \right) \ dV 
	+ O\left(\frac{1}{R_1^2}\right) .
\end{equation}
Combining Eqs.~(\ref{VolInt_approx1}) and (\ref{I_2_approx}) 
we arrive at the final result 
\begin{equation}
\label{VolInt_approx2}
\int \nabla^2 u \ dV =  
I_1 + \int_{R_1}^{R_2} f\left( -\frac{1}{4\pi}\frac{I_1}{r}  \right) \ dV
    + O\left(\frac{1}{R_1^2}\right) ,
\end{equation}
which depends only on $I_1$ computed from $u$ on the numerical grid and
a volume integral outside the grid over the known function $f$.

We have applied the method described in the previous paragraph to both the
ADM and Komar mass integrals at infinity, as both can be expressed as volume
integrals. Fig.~\ref{M_ADM_grid_extension} shows $M^{ADM}_{\infty}$
computed using the extended volume integral
(\ref{VolInt_approx2}) as well as the result if we integrate only over the
grid using Eq.~(\ref{VolInt_approx0}). The extended volume integral result
is clearly more accurate, say for $R_1 =12M$, which we have used to obtain
our sequence.

\vfill
\mbox{}

\bibliography{references}

\end{document}